\documentclass[11pt, a4paper,english,superscriptaddress,aps]{revtex4}
\usepackage{amsmath,amssymb,graphicx}
\makeatletter

\usepackage{babel}
\usepackage[active]{srcltx}
\usepackage{graphicx,color}
\usepackage{changebar}
\usepackage{hyperref}
\usepackage[T1]{fontenc}
\usepackage{ae}
\usepackage[ansinew]{inputenc}
\usepackage{amsmath}
\usepackage{braket}
\usepackage{mathtools}
\usepackage{slashed}
\usepackage{empheq}
\usepackage{comment}

\usepackage{graphicx}
\usepackage{amsfonts}
\usepackage{amsmath}
\def\m{\mu}

\newcommand{\be}{\begin{equation}}
\newcommand{\ee}{\end{equation}}
\newcommand{\bea}{\begin{eqnarray}}
\newcommand{\eea}{\end{eqnarray}}

\begin{document}

\title{One-loop corrections in Maxwell-metric-affine bumblebee gravity}

\author{A. C. Lehum}
\email[]{lehum@ufpa.br}
\affiliation{Faculdade de F\'{i}sica, Universidade Federal do Par\'{a}, 66075-110, Bel\'{e}m, Par\'a, Brazil}

\author{J. R. Nascimento}
\email[]{jroberto@fisica.ufpb.br}
\affiliation{Departamento de F\'{\i}sica, Universidade Federal da 
Para\'{\i}ba\\
 Caixa Postal 5008, 58051-970, Jo\~ao Pessoa, Para\'{\i}ba, Brazil}
 
\author{A. Yu. Petrov}
\email[]{petrov@fisica.ufpb.br}
\affiliation{Departamento de F\'{\i}sica, Universidade Federal da 
Para\'{\i}ba\\
 Caixa Postal 5008, 58051-970, Jo\~ao Pessoa, Para\'{\i}ba, Brazil}

\author{P. J. Porf\'{i}rio}\email[]{pporfirio@fisica.ufpb.br}
\affiliation{Departamento de F\'{\i}sica, Universidade Federal da Para\'{\i}ba\\
 Caixa Postal 5008, 58051-970, Jo\~ao Pessoa, Para\'{\i}ba, Brazil}

\begin{abstract}
In this paper, we consider the coupling of the metric-affine bumblebee gravity to the Abelian gauge field and obtain the effective model corresponding to the weak gravity limit of this theory. The effective bumblebee theory displays new unconventional couplings between the bumblebee field and its field strength, and the $U(1)$ gauge field along with its respective field strength, as a result of the non-metricity effects. Thus, being a new gauge-bumblebee theory, it represents an example of vector-vector couplings which are very rarely considered, if not entirely overlooked, in the Abelian case. For this theory we calculate the lower perturbative corrections. We close the paper with discussions of other possible vector-vector couplings.
\end{abstract}

\maketitle

\section{Introduction}

As it is well known, spontaneous symmetry breaking mechanism is one of the most powerful ways to violate Lorentz symmetry within various field theories, especially in gravity (for discussions, see f.e. \cite{KosLiGrav} and references therein). Indeed, within this approach one does not need to introduce an {\it a priori} constant vector (it must be noted that, in general, constant vectors cannot be consistently defined in a curved spacetime \cite{KosLiGrav}), but, instead of this, the Lorentz symmetry is broken through arising of some non-trivial vacuum of a bumblebee potential \cite{KosGra}. We note that namely this mechanism was assumed originally in \cite{KosSam} to introduce a possibility for the Lorentz symmetry breaking (LSB) itself. It is interesting not only in a curved spacetime but also in a flat one as well. Therefore, the natural task is to consider more generic field theory models involving different fields coupled to a bumblebee field, both at classical and quantum levels.

Earlier, one of the main interests within the LSB context was called by the bumblebee-spinor models in flat spacetime allowing for dynamical LSB, so that the one-loop effective potential arises as a fermionic determinant \cite{dyn1,dyn2,dyn3}. Another pivotal topic we need to bring to bear is the incorporation of (local) LSB in the context of gravitational theories. Fortunately, a consistent way to implement LSB in curved space-times has already been developed in \cite{KosGra}. In this work, the authors propose a general framework laying out all possible Lorentz-violating operators in the pure-gravity sector of the Standard Model Extension (SME) which is defined in a Riemann-Cartan manifold. Furthermore, they present the bumblebee gravity model as an example of a modified theory of gravity with spontaneous LSB. Although such a model was originally scrutinized in the Riemann-Cartan background, a flurry of works has addressed it in the standard pseudo-Riemannian background \cite{C,D,E,F,G,H,L,M}. However, as it is well known, taking a more generic geometrical approach than the usual metric one, we mean considering non-Riemannian geometries, might be a promising way to explore further theoretical and phenomenological consequences of the LSB within modified theories of gravity.  To go forward in this subject, the authors of \cite{ourbumb2019} proposed a metric-affine extension of the bumblebee gravity model called metric-affine bumblebee gravity. From the theoretical perspective, they have obtained new unusual vector-spinor and vector-scalar couplings to arise within this model  \cite{ourbumb2019}. These theories, afterwards, have been treated perturbatively \cite{ourbumb2,ourbumb3}, and some studies of perturbative aspects of the bumblebee field coupled to other fields are presented also in \cite{Maluf1, Maluf2}. From the phenomenological perspective, exact solutions describing rotating and non-rotating black holes with LSB have been obtained within this model \cite{Filho:2022yrk, Filho:2024hri}, and the upper bound limits of the Lorentz-violating (LV) parameter have been estimated by confronting the theoretical results with the astrophysics experimental data at disposal \cite{Filho:2022yrk, Filho:2024hri}.

Therefore, due to the richness of the metric-affine bumblebee gravity model, the natural continuation of the aforementioned studies consists in coupling our model to a gauge field. The importance of studying such a modified theory arises in different scenarios. First, in the context of astrophysics, exact (non-)rotating charged black hole solutions play an important role since they can give us constraints on our model by confronting the theoretical predictions with the current data or at least bring signals of new physics beyond GR. Second, in the cosmological scenario, the three-point and four-point correlation functions of scalar, vectorial and tensorial quantities can generate new physical effects by comparing with the available experimental data, thus it should be an important route to be investigated. Nonetheless, such a study requires a detailed construction of background inflationary models and then its correspondent perturbation theory, this by itself is a challenging research program and should be studied elsewhere. Besides this, we note that study of electrodynamics in a curved background has been performed, within the Riemannian gravity framework, in many papers, being motivated by the problem of primordial electromagnetic fields (see e.g. \cite{Tsagas,Tsagas2}) and many other related issues, including use of the electromagnetic field as a source allowing to obtain many nontrivial gravitational solutions beginning from the paradigmatic Reissner-Nordstr\"{o}m one, and including for example trigonometric solutions for G\"{o}del-type metrics \cite{RT} and many other cases. Therefore, the generalization of these studies to the metric-affine gravity, in particular, studying of perturbative aspects of such a theory, which can be especially important in the primordial Universe,  seems to be a very interesting problem. We note that namely in the earlier epoch of the Universe the impact of Lorentz symmetry breaking could be essentially important, and since the most natural manner to introduce the LSB in gravity is based on the bumblebee framework \cite{KosGra}, studies of the electrodynamcs coupled to the bumblebee gravity appear to be a very natural task.
In this work, we shall focus on the weak field regime, in other words, we are intended to introduce the low-energy effective gauge-gravity theory. In this paper, we pursue this aim.  Explicitly, we formulate a theory where the metric-affine bumblebee gravity is coupled to a gauge field, consider the weak gravity limit of this theory, obtaining, as a result, the effective bumblebee-gauge theory, and calculate lower quantum corrections in it.

The structure of this paper looks like as follows. In section II, we briefly review the main aspects of the metric-affine bumblebee gravity and formulate our model. In section III, we proceed with the quantum corrections of the effective metric-affine bumblebee model coupled with a gauge field in the weak field approximation and expanded up to the second order in the non-minimal coupling. In section IV, we summarize our results.

\section{Bumblebee model in Palatini approach}

We start our study by writing the bumblebee action in a curved space-time (cf. f.e. \cite{Seifert}):
\begin{eqnarray}
S_{B}&=&\int d^4 x\,\sqrt{-g}\Big[\frac{1}{2\kappa^2}\Big(R(\Gamma)+\xi B^{\alpha} B^{\beta} R_{\alpha\beta}(\Gamma)\Big)-\frac{1}{4}B^{\mu\nu}B_{\mu\nu}-V(B^{\mu}B_{\mu}\mp v^2)\Big] + \nonumber\\ 
&+& \int d^4 x\,\sqrt{-g}\mathcal{L}_{M}(g_{\mu\nu},A_{\nu}),
\label{bumblebee}
\end{eqnarray} 
where $\sqrt{-g}$ is the determinant of the metric $g_{\mu\nu}$, $\mathcal{L}_{M}(g_{\mu\nu},A_{\nu})$ is the Lagrangian involving all non-gravitational fields, in our case, the electromagnetic field, and $\xi$ is the constant characterizing the non-minimal coupling between the bumblebee field and geometry of space-time. Here we assume that the gravitational sector of (\ref{bumblebee}) is defined in the metric-affine approach in which the connection $\Gamma^{\mu}_{\nu\lambda}$  and metric $g_{\mu\nu}$ are independent dynamical fields {\it a priori}. 

Another ingredient is the bumblebee field $B_{\mu}$ introduced in order to break the Lorentz symmetry spontaneously \cite{KosGra}. We remark on the presence of the non-minimal coupling between the bumblebee and the Ricci tensor (the second term in the r.h.s. of (\ref{bumblebee})) which will play an important role in the following.
  The potential $V(B^{\mu}B_{\mu}\mp v^2)$ is chosen in such a way that the bumblebee field $B_{\mu}$ acquires the non-zero expectation value (VEV) given by some vector whose presence introduces the privileged direction in the space-time, hence resulting in a spontaneous Lorentz symmetry breaking.   Explicitly, we use the most natural form for this potential, that is, $V=\frac{\lambda}{4}(B^{\mu}B_{\mu}\mp v^2)^2$, here and further, the $\mp$ sign reflects the fact that the minimum of $B_{\mu}$ can be both {time}-like and {space}-like, while $v^2>0$. Further, we shall denote the bumblebee field strength associated with $B_{\mu}$ by $B_{\mu\nu}$, and its explicit form is defined below
\begin{eqnarray}
B_{\mu\nu}&\equiv&\ (dB)_{\mu\nu}
=
\partial_{\mu}B_{\nu}-\partial_{\nu}B_{\mu}.
\end{eqnarray}
Here, for the sake of the simplicity, we assumed the zero torsion case similarly to the previous papers \cite{ourbumb2019,ourbumb2}, hence, the affine connection is symmetric. We note that, despite the kinetic term of the bumblebee field is similar to that one of the electromagnetic field, the bumblebee field itself does not possess the gauge symmetry because of the nontrivial potential. The electromagnetic field Lagrangian is assumed to be the usual one
\begin{equation}
	\label{lagrem}
{\cal L}_A=-\frac{1}{4}\sqrt{-g}F_{\mu\nu}F^{\mu\nu}.
\end{equation}

The equations of motion in our theory, for a generic external fields (their role in our case is played by the electromagnetic field) look like \cite{ourbumb2}:
	\begin{eqnarray}
		\nonumber 0&=&R_{(\mu\nu)}(\Gamma)-\frac{1}{2}g_{\mu\nu}\bigg(R(\Gamma)+\xi B^{\alpha}B^{\beta}R_{\alpha\beta}(\Gamma)\bigg)+
		+2\xi\bigg(B_{(\mu}R_{\nu)\beta}(\Gamma)\bigg)B^{\beta}=\kappa^2  T_{\mu\nu},\label{Riccieq}\nonumber\\
		0&=&\nabla_{\lambda}^{(\Gamma)}\bigg[\sqrt{-g}g^{\mu\alpha}\bigg(\delta^{\nu}_{\alpha}+\xi B^{\nu}B_{\alpha}\bigg)\bigg],\label{connectioneq}\nonumber\\
		\nabla_{\mu}^{(g)}B^{\mu\nu}&=&-\frac{\xi}{\kappa^2}g^{\nu\alpha}B^{\beta}R_{\alpha\beta}(\Gamma)+2 V^{\prime}B^{\nu},\label{bumblebeeeq}
		\label{PDE}
	\end{eqnarray}
	where $T_{\mu\nu}=T_{\mu\nu}^{M}+T_{\mu\nu}^{B}$ described the contributions of the matter sources $T_{\mu\nu}^{M}$ (in our case this is just the energy-momentum tensor of the electromagnetic field) and of the bumblebee field, namely
	\begin{eqnarray}
		T_{\mu\nu}^{M}&=&-\frac{2}{\sqrt{-g}}\frac{\delta(\sqrt{-g}\mathcal{L}_{M})}{\delta g^{\mu\nu}},\\
		T_{\mu\nu}^{B}&=& B_{\mu\sigma}B_{\nu}^{\ \sigma}-\frac{1}{4}g_{\mu\nu}B^{\alpha}_{\ \sigma}B^{\sigma}_{\ \alpha}-V g_{\mu\nu}+2V^{\prime}B_{\mu}B_{\nu}.
	\end{eqnarray}
	The equation of motion for the connection (the second one in (\ref{PDE})) in our case is the purely algebraic one, its solution is given by the Levi-Civita connection of the auxiliary metric $h_{\mu\nu}$ defined by
	\begin{equation}
		h_{\mu\nu}=\sqrt{1+\xi X}\left(g_{\mu\nu}-\frac{\xi}{1+\xi X} B_{\mu}B_{\nu}\right),\label{hmn}
	\end{equation}  
 and, its inverse,
 \begin{equation}
		h^{\mu\nu}=\frac{1}{\sqrt{1+\xi X}}\left(g^{\mu\nu}+\xi B^{\mu}B^{\nu}\right),\label{Hmn}
	\end{equation}  
	where $X \equiv g^{\mu\nu}B_{\mu}B_{\nu}$.  Here, the quantity $X$ should be understood as a function of $Y\equiv h^{\mu\nu}B_{\mu}B_{\nu}$. Since $\nabla_{\lambda}^{(\Gamma)}h_{\mu\alpha}=0$, the nonmetricity tensor, $Q_{\lambda\mu\alpha}=\nabla_{\lambda}^{(\Gamma)}g_{\mu\alpha}$, takes the explicit form 
	\begin{equation}
	Q_{\lambda\mu\alpha}=\frac{\xi}{1+\xi}\left(B_{\mu}\nabla_{\lambda}^{(\Gamma)}B_{\alpha}+B_{\alpha}\nabla_{\lambda}^{(\Gamma)}B_{\mu}\right).
	\end{equation}
	Thus the nonmetricity tensor is completely sourced by the bumblebee. With the help of Eqs.\eqref{hmn} and \eqref{Hmn}, we can integrate out $g_{\m\nu}$ of the action \eqref{bumblebee} and then obtain its Einstein-frame representation, whose explicit form reads
	\begin{equation}
	\label{eframe}\tilde{S}_{B}=\int d^4 x\,\sqrt{-h}\frac{1}{2\kappa^2}R(h)+\int d^4 x\,\sqrt{-h}\tilde{\mathcal{L}}_{M}(h_{\mu\nu},B_{\nu},A_{\nu}),
	\end{equation}
	where
	\begin{eqnarray}
	\nonumber\tilde{\mathcal{L}}_{M}(h_{\mu\nu},B_{\nu},A_{\nu})&=&\mathcal{L}_{M}(g^{\mu\nu}(h_{\mu\nu},X,B_{\nu}),B_{\nu},A_{\nu})-\frac{1}{4}\tilde{B}^{\mu\nu}B_{\mu\nu}- V(\tilde{B}^{\mu}B_{\mu}\mp v^2)+\\
	&+&\tilde{V}(B_{\mu}, X, h_{\mu\nu}),
	\end{eqnarray}
where contravariant tilded tensorial quantities denote that their indices were raised using $h^{\mu\nu}$ {and $\tilde{V}$ is a function of $B_{\mu}, X$ and $ h_{\mu\nu}$}. Note that the gravitational sector, in the Einstein-frame representation, reduces to GR (of the auxiliary metric), while the matter sector is modified. It happens because the non-minimal interaction between the bumblebee and the connection, in the gravity sector of \eqref{bumblebee}, is now shifted to the new modified matter sector in this representation (\ref{eframe}),  which in turn involves unconventional vector-vector interactions. As pointed out in \cite{ourbumb2019}, the Einstein-frame representation is useful for investigating the weak field regime,  which means neglecting the Newtonian and post-Newtonian effects (turning off the curvature effects) and only maintaining the nonmetricity ones. 

Now, in order to proceed with a perturbative description of the theory, we assume particle physics scenarios that correspond to the weak gravitational field regime. In this context, the auxiliary metric is $h_{\mu\nu}=\eta_{\mu\nu}$ and $\xi\ll 1$. On the other hand, the physical (spacetime) metric $g_{\mu\nu}$ gets local contributions of the energy density stemming from the bumblebee field (contributions of the nonmetricity tensor). To see this in more detail, let us invert \eqref{hmn}, we have 
\begin{equation}
    g_{\mu\nu}=\frac{1}{\sqrt{1+\xi X}}h_{\mu\nu}+\frac{\xi}{1+\xi X}B_{\mu}B_{\nu}.
\end{equation}
In this scenario, applying the weak field regime in this equation, one has
\begin{equation}
    g_{\mu\nu}=\eta_{\mu\nu}\left(1-\frac{1}{2}\xi X+\frac{3}{8}\xi^2 X^2\right)+\xi(1-\xi X)B_{\mu}B_{\nu}.
    \label{eq}
\end{equation}
It is worth recalling that $X=g^{\mu\nu}B_{\mu}B_{\nu}$ and $Y=h^{\mu\nu}B_{\mu}B_{\nu}$. They are related as $Y=X\sqrt{1+\xi X}$, where we have used Eq.\eqref{Hmn}. Then, in the weak field regime, after some algebraic manipulation, we find the following relations: $\xi Y=\xi B^2=\xi \eta^{\mu\nu}B_{\mu}B_{\nu}$ and $\xi X=\xi B^2-\frac{1}{2}\xi^2 B^4$. Substituting them into Eq.\eqref{eq}, one finds
\begin{equation}
  g_{\mu\nu}=\eta_{\mu\nu}+\xi(B_{\mu}B_{\nu}-\frac{1}{2}B^2\eta_{\mu\nu})+\xi^2 B^2\left(\frac{5}{8}\eta_{\mu\nu}B^2-B_{\mu}B_{\nu}\right),
  \label{gexpand}
  \end{equation}
  and its inverse
  \begin{equation}
  g^{\mu\nu}=\eta^{\mu\nu}+\xi(\frac{1}{2}B^2\eta^{\mu\nu}-B^{\mu}B^{\nu})+\xi^2 B^2\left(B^{\mu}B^{\nu}-\frac{3}{8}\eta^{\mu\nu}B^2\right),
  \label{gexpand1}
  \end{equation}
  where $B^2 \equiv \eta^{\mu\nu}B_{\mu}B_{\nu}$. It is straightforward from Eq.\eqref{gexpand} that there is a residual contribution(even turning off the Newtonian and post-Newtonian effects) responsible for the local departure from the Minkowski metric  (we call the post-Minkowskian limit). Therefore, in this limit, the effective theory turns out to yield new interaction terms, as a consequence of non-trivial nonmetricity effects. {From now on, indices are raised and lowered by $\eta^{\mu\nu}$ and $\eta_{\mu\nu}$, respectively. Let us consider our particular model in the post-Minkowskian regime}. Substituting Eqs. \eqref{gexpand} and \eqref{gexpand1} in the electromagnetic Lagrangian (\ref{lagrem}), we find	
\begin{eqnarray}
\nonumber {\cal L}_A&=&-\frac{1}{4}F_{\mu\nu}F^{\mu\nu}-\frac{1}{8}\xi B^2F_{\mu\nu}F^{\mu\nu}+\frac{\xi}{2}B^{\mu}F_{\mu\nu}B_{\lambda}F^{\lambda\nu}+\frac{3}{32}\xi^2B^4 F^{\mu\nu}F_{\mu\nu}-\\
&-&\frac{\xi^2}{2}B^2 F^{\mu\beta}F_{\mu\nu}B^{\nu}B_{\beta}+O(\xi^3).
\label{Photon}
\end{eqnarray}
We note that the third term in this expression is nothing more than the aether term \cite{CarTam,ouraether}.
This action should be summed with the bumblebee action \eqref{bumblebee} in the post-Minkowskian limit \cite{ourbumb2} and expanded up to second order in $\xi$ by using Eqs. (\ref{gexpand},\ref{gexpand1}), thus
\begin{eqnarray}
\nonumber {\cal L}_{BEF}&=&-\frac{1}{4}B_{\mu\nu}B^{\mu\nu}-\frac{\lambda}{4}\left(B^2- v^2\right)^2\\ 
\nonumber 
&+&\frac{\xi}{4}\left[\frac{\lambda}{2}\left(B^2 v^4-4B^4 v^2+3B^6\right) -\frac{1}{2} B^2 B_{\mu\nu}B^{\mu\nu}+2B^{\mu}_{\phantom{a}\beta}B_{\mu\nu}B^{\nu}B^{\beta}\right]\\
\nonumber
&+&\frac{\xi^2}{8}\bigg[\frac{\lambda}{4}(-5B^4 v^4 + 24 B^6 v^2 - 21 B^8)+\frac{3}{4}B^4 B_{\mu\nu}B^{\mu\nu}-\\
&-&4B^2 B^{\mu\beta}B_{\mu\nu}B^{\nu}B_{\beta}\bigg]+\mathcal{O}(\xi^3),
	\label{BumbLagPert}
\end{eqnarray}	
{where we have chosen the $(-)$ sign to have a time-like vector for the minimum of $B_{\mu}$.}

  It should be realized that the above Lagrangian is more generic than the one considered in \cite{ourbumb2}, where terms proportional to $\xi^2$ and $\xi\lambda$ were disregarded. So, we can study various aspects of the theory formed by the sum ${\cal L}_A+{\cal L}_{BEF}$, for example, exact solutions and quantum corrections.

Formally, due to post-Minkowskian corrections, which stem from the non-metricity effects, the bumblebee potential takes the following form
\begin{eqnarray}
    \nonumber V(B^2)&=&\frac{\lambda}{4}\left(B^2- v^2\right)^2-\frac{\xi}{4}\left[\frac{\lambda}{2}\left(B^2 v^4-4B^4 v^2+3B^6\right)\right]\\
&-&\frac{\xi^2}{8}\bigg[\frac{\lambda}{4}(-5B^4 v^4 + 24 B^6 v^2 - 21 B^8)\bigg]+\mathcal{O}(\xi^3),
	\label{Potential}	
\end{eqnarray}
which is responsible for spontaneous LSB. The minimum of the {former} potential  must satisfy
\begin{equation}
    \frac{d V}{d B}\bigg|_{B_{\mu}=b_{\mu}}=0,
\end{equation}
whose solution, up to second-order in $\xi$, occurs at
\begin{equation}
b^2=v^2+\xi\frac{v^4}{4}-\xi^2 \frac{v^6}{8},
\label{bb}
\end{equation}
where $b^2=\eta^{\mu\nu}b_{\mu}b_{\nu}$ and $\langle B^\mu \rangle = b^\mu$  is the nonzero VEV. Therefore, in practical terms, post-Minkowskian corrections are responsible for the departures from the standard condition of the minimum of the potential in the metric approach, that is, $b^2=v^2$ (we note that at this step, some estimations for the LV parameters can be done. Indeed, it was argued in \cite{Filho:2022yrk, Filho:2024hri}, that it follows from advance of the Mercury perihelion that the dimensionless LV parameter $X=\xi b^2<4.9\cdot 10^{-12}$). However, it is noteworthy that the authors of \cite{ourbumb2} chose, for simplicity, to neglect the post-Minkowskian corrections to the condition \eqref{bb}. To study the quantum effects of the model, it is necessary to shift the bumblebee field as $B^\mu \rightarrow B^\mu + b^\mu$ so that the new quantum field has a vanishing VEV. Proceeding with this, the effective Lagrangian \eqref{BumbLagPert} becomes
\begin{eqnarray}
	{\cal L}_{BEF}&=&-\frac{1}{4}\left[\eta_{\nu\beta}\left(1+\frac{\xi}{2}b^2-\frac{3}{8}\xi^2 b^4\right)-2\xi b_{\beta}b_{\nu}+\frac{1}{2}\xi^2 b^2 b_{\beta}b_{\nu}\right]B^{\mu\nu}B_{\mu}^{\phantom{a}\beta}-\nonumber\\
 \nonumber&-&\frac{\xi}{8}B^2 B^{\mu\nu}B_{\mu\nu}-\frac{\xi}{4}(B\cdot b)B_{\mu\nu}B^{\mu\nu}+ \frac{\xi}{2}B^{\nu}B^{\beta}B^{\mu}_{\phantom{a}\beta}B_{\mu\nu}+\xi B^{\mu}_{\phantom{a}\beta}B_{\mu\nu}B^{\nu}b^{\beta}+\\
 \nonumber&+&\frac{3}{32}\xi^2B^{\mu\nu}B_{\mu\nu}\left[B^4+2B^2 b^2+4(B\cdot b)(B^2 +b^2)+4(B\cdot b)^2\right]-\\
 \nonumber&-&\frac{\xi^2}{2}(B^2+2B\cdot b +b^2)B^{\mu\beta}B_{\mu\nu}B^{\nu}B_{\beta}-\xi^2(B^2+2B\cdot b +b^2)B^{\mu\beta}B_{\mu\nu}B^{\nu}b_{\beta}-\\
 \nonumber&-&\frac{\xi^2}{2}(B^2+2B\cdot b)B^{\mu\beta}B_{\mu\nu}b^{\nu}b_{\beta}-\\
\nonumber &-&\frac{1}{4}\lambda(B^2 +b^2 +2B\cdot b -v^2)^2+\\
 \nonumber&+&\frac{1}{8}\xi\lambda\Big[v^4(B^2 + b^2 +2B\cdot b)-4v^2(B^2 + b^2 +2B\cdot b)^2 + 3(B^2 + b^2 +2B\cdot b)^3\Big]-\\
\nonumber &-&\frac{1}{32}\xi^2 \lambda\Big[5v^4(B^2 + b^2 +2B\cdot b)^2 - \nonumber\\
&-& 24v^2(B^2 + b^2 +2B\cdot b)^3+21(B^2 + b^2 +2B\cdot b)^4\Big]+\nonumber\\
&+&\mathcal{O}(\xi^3).\label{BumbLagPert1}
\end{eqnarray}
With this and using Eq.\eqref{bb}, we can find the mass-squared tensor for the effective theory by defining
\begin{eqnarray}
    M_{\alpha\beta}=-\frac{d^2 {\cal L}_{BEF}}{dB^{\alpha} dB^{\beta}}\bigg|_{B^{\mu}=0}=\frac{d^2 V_{eff}}{dB^{\alpha} dB^{\beta}}\bigg|_{B^{\mu}=0},
\end{eqnarray}
whose explicit form is given by
\begin{equation}
    M_{\alpha\beta}=-2\lambda b_{\alpha}b_{\beta}-5\xi\lambda v^2 b_{\alpha}b_{\beta}+\frac{41}{4}\xi^2 \lambda v^4 b_{\alpha}b_{\beta} +\mathcal{O}(\xi^3).
\end{equation}

The next step is to determine the propagators. In \cite{ourbumb2}, a non-massive bumblebee propagator was found up to first order in $\xi$, indicating the absence of a massive pole. This enables us to adopt a different approach from that used in \cite{ourbumb2}. Specifically, we will consider the free propagators for the bumblebee and gauge fields, while treating their couplings to the background field ($b_{\alpha}$) as perturbative interactions. These couplings will appear in the Feynman diagrams as insertions.


In addition, it is necessary to include a longitudinal term for the bumblebee field, $\mathcal{L}_{\zeta_b} = \frac{1}{2\zeta_b}(\partial^\mu B_{\mu})^2$, with $\zeta_b$ being the longitudinal gauge coupling, to facilitate the calculation of the bumblebee propagator. Indeed, if we consider the action (\ref{BumbLagPert1}) at $\xi=0$, we recover, in the bumblebee sector, just the standard Maxwell Lagrangian $-\frac{1}{4}B_{\mu\nu}B^{\mu\nu}$. Therefore, the propagator of the bumblebee field will be analogous to that one for the electromagnetic field, while all terms proportional to the $b_{\mu}$ LV vector and all couplings, as we already said, will be treated as perturbations. We note that the $B_{\mu}$ field, being the gauge one, is characterized just by two degrees of freedom which is natural since it is straightforwardly related with the metric $g_{\mu\nu}$, which, as it is well known, is described just by two degrees of freedom, through the relation (\ref{gexpand}).

Thus, the propagators are the free ones looking like
 \begin{eqnarray}
     \nonumber <B^{\mu}(-p)B^{\nu}(p)>&=&\frac{i}{p^2}(\eta^{\mu\nu}-(1-\zeta_b)\frac{p^{\mu}p^{\nu}}{p^2});\\
     <A^{\mu}(-p)A^{\nu}(p)>&=&\frac{i}{p^2}(\eta^{\mu\nu}-(1-\alpha)\frac{p^{\mu}p^{\nu}}{p^2}).
 \end{eqnarray}
 Here, $\alpha$ is the gauge parameter corresponding to the usual gauge fixing term
 \begin{equation}
     {\cal L}_{gf}=\frac{1}{2\alpha}\left(\partial_{\mu}A^{\mu}\right)^2,
 \end{equation}
\noindent while the vertices can be read off from Eqs. \eqref{Photon} and \eqref{BumbLagPert} by using the Feynman rules. Since we are dealing with an Abelian gauge theory, the contributions from the Faddeev-Popov ghosts are trivial.
 
\section{Radiative corrections in the LV broken phase}

Now, let us calculate some perturbative contributions in the effective theory. Initially, we focus on deriving the contribution involving external gauge legs, with the bumblebee field is integrated out, and further, we obtain the bumblebee-dependent corrections. To study the lower quantum contributions in our theory, we restrict our calculation to the order of $\xi^2$ and $b^2$.

\subsection{The photon self-energy}

Let us begin the computation of the one-loop effective Lagrangian for the photon self-energy,
\begin{eqnarray}
	\mathcal{L}_{eff}= \mathcal{L}_{eff}^{(0)} + \mathcal{L}_{eff}^{(1)} + \cdots,
\end{eqnarray}
\noindent where the superscripts $(0)$ and $(1)$ denote the tree-level and one-loop order contributions, respectively.

After performing the shift $B^\mu \rightarrow B^\mu + b^\mu$, we can rewrite the tree-level quadratic part of the photon field effective Lagrangian as
\begin{eqnarray}
\label{treelevel}
 \mathcal{L}_{eff}^{(0)} &=&-\frac{(1+\delta_3)}{4}F_{\mu\nu}F^{\mu\nu}
	-\frac{(\xi+\delta_4)}{8} b^2 F_{\mu\nu}F^{\mu\nu} +\frac{(\xi+\delta_4)}{2} b^{\mu} F_{\mu\nu} b_{\lambda} F^{\lambda\nu}\nonumber\\
&&+\mathcal{O}(b^4)+\mathrm{interactions},
\end{eqnarray}
\noindent where $\delta_3$ and $\delta_4$ are the counterterms associated with their respective operators.

\begin{figure}[ht]
	\includegraphics[angle=0 ,width=14cm]{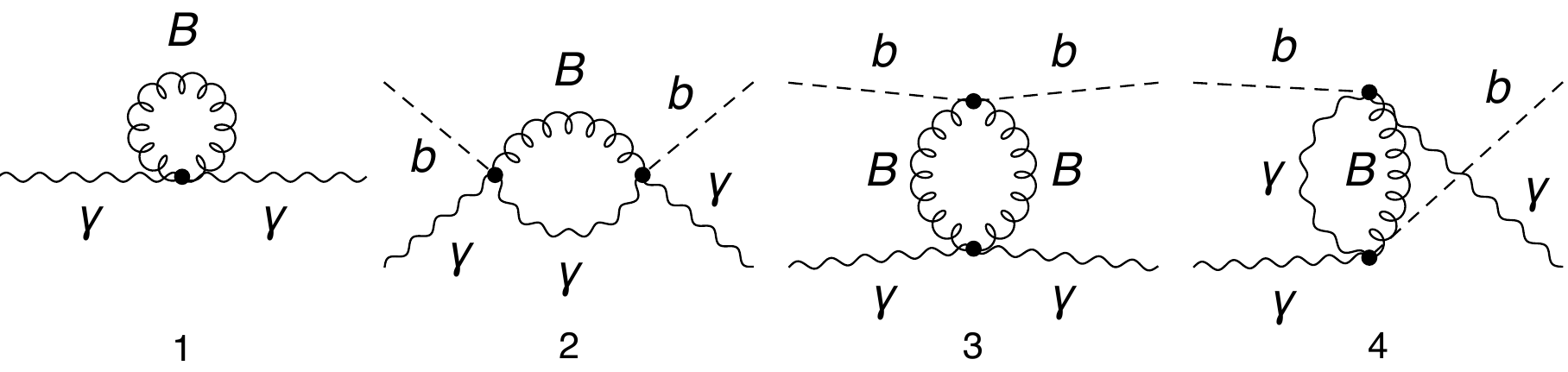}
	\caption{One-loop photon self-energy. Wavy and curly lines denote photon and Bumblebee propagators, respectively, while dashed lines signify the insertion of the Lorentz-violating field $b^{\mu}$. }
	\label{fig01}
\end{figure}

Next, we turn to compute the radiative corrections to the photon self-energy. The diagrams contributing to this process are illustrated in Figure \ref{fig01}. The expression for diagram \ref{fig01}.1 is proportional to a single loop of a massless particle, resulting in a vanishing contribution. Therefore, we proceed to calculate the effects of the insertion of the LV vertices. The corresponding diagrams are depicted in Figure \ref{fig01}.2-4. The expression for the polarization tensor can be written as
\begin{eqnarray}\label{sephoton1}
	-i(2\pi)^D \Pi^{\mu\nu}_A(p)  &=& \left[g^{\mu \nu } (b^2 p^2- 2(b\cdot p)^{2})+p^\mu(2b^\nu (b\cdot p)-b^2 p^\nu)+2 b^\mu (p^\nu (b\cdot p)-p^2b^\nu )\right]\nonumber\\
	&&\times  \frac{\pi^2 \xi^2 }{6}p^2\Big[ 
		p^2(\zeta_b-1)\textbf{C}_0\left(0,p^2,p^2,0,0,0\right)
	-2(\zeta_b-2) \textbf{B}_0\left(p^2,0,0\right)\Big]\nonumber\\
	&&+\frac{\pi^2 \xi^2}{6}\left(p^2g^{\mu\nu}-p^\mu p^\nu \right) \left(2 (b \cdot p)^2-b^2 p^2\right)\Big[ 2(1+\zeta_b)  \textbf{B}_0\left(p^2,0,0\right)\nonumber\\
	&&- p^2 (\zeta_b-1)  \textbf{C}_0\left(0,p^2,p^2,0,0,0\right)\Big],
\end{eqnarray}

\noindent where $D=4-2\epsilon$ is the dimension of the spacetime. The integrals $\textbf{B}_0$ and $ \textbf{C}_0$ are defined in the appendix. We are using dimensional regularization (DR) and minimal subtraction (MS) as renormalization scheme. We adopt the prescription where all scaleless integrals vanish \cite{Abreu:2022mfk}, specifically $\textbf{B}_0(0,0,0)=0$.

The one-loop correction to the photon self-energy leads to the following nonlocal contribution to the quadratic part of the photon effective Lagrangian:
\begin{eqnarray}
\label{oneloop}
	\mathcal{L}_{eff}^{(1)} &=&
	-\frac{1}{4}F_{\mu\nu}\left[-\frac{i\pi^2 \xi ^2}{6(2\pi)^D} (3\zeta_b+1)\textbf{B}_0\left(p^2,0,0\right) \left(2 (b\cdot \partial)^2-b^2 \Box\right)\right] F^{\mu\nu}\nonumber\\
	&&-\frac{1}{8} b^2F_{\mu\nu} \left[-\frac{i\pi^2 \xi^2}{6(2\pi)^D} (5-3\zeta_b)\textbf{B}_0\left(p^2,0,0\right)\Box \right] F^{\mu\nu} 
	\nonumber\\
	&& + \frac{1}{2}b^{\mu}F_{\mu\nu} \left[-\frac{i\pi^2 \xi^2}{6(2\pi)^D} (5-3\zeta_b)\textbf{B}_0\left(p^2,0,0\right)\Box \right] b_{\lambda}F^{\lambda\nu}. 
\end{eqnarray} 
\noindent where we have used Tarasov's algorithmic approach \cite{Tarasov:1997kx} to express the integral $\textbf{C}_0$ in terms of $\textbf{B}_0$ (see the Appendix).

As we can see, the one-loop effective Lagrangian for the photon field exhibit a LV Podolsky-like term, followed by non-local aether-like terms. The nonlocality in the one-loop effective Lagrangian arises from the massless bumblebee and photon fields in the internal loop (see Fig. \ref{fig01}), stemming from the long-range nature of interactions mediated by massless particles. The UV divergences must be absorbed by the same operators.

It is interesting now to discuss the possible modification of the static potential of the electromagnetic field caused by new additive LV terms. The one-loop corrected effective Lagrangian of the gauge field, given by the sum of (\ref{treelevel}), (\ref{oneloop}), and the standard source-dependent term,  can be written as
\bea
{\cal L}_{eff}=-\frac{1}{4}F_{\mu\nu}\left[1+a_1\xi^2[2\tilde{a}_1(b\cdot\partial)^2-a_1b^2\Box]
\right]F^{\mu\nu}+\frac{1}{2}a_2\xi^2b^{\mu}b_{\lambda}F_{\mu\nu}\Box F^{\lambda\nu}+j_{\mu}A^{\mu},
\eea
where $\tilde{a}_1,a_1,a_2$ are some numbers.
The corresponding equations of motion look like
\bea
\left[1+\xi^2[2\tilde{a}_1(b\cdot\partial)^2-a_1b^2\Box]
\right]\partial_{\mu}F^{\mu\nu}-a_2\xi^2\left[
(b\cdot\partial)\Box (b_{\lambda}F^{\lambda\nu})-b^{\nu}b_{\lambda}\partial_{\mu}\Box F^{\lambda\mu}
\right]=j^{\nu}.
\eea
Let us consider the static potential solutions for this equation. To do it, we choose $A^{\mu}(x)=(\Phi(\vec{x}),\vec{0})$, $j^{\mu}=(\rho(\vec{x}),\vec{0})$, hence the only non-zero components of $F^{\mu\nu}$ are $F^{0i}=-F^{i0}=-\partial^i\Phi$. Also, for the sake of the simplicity, we choose the LV vector to be $b^{\mu}=(b,0,0,0)$, to be directed along the time axis. As we restrict ourselves to the static solution, we have $(b\cdot\partial)\Phi=0$ for this case, and $\Box\to -\nabla^2$, and our equation is simplified:
\bea
\left[1+a_1\xi^2b^2\nabla^2
\right]\nabla^2\Phi-a_2\xi^2b^2(\nabla^2)^2\Phi=\rho.
\eea
This equation can be solved with the Green function method: we choose $\Phi(\vec{x})=\int d^3\vec{y}G(\vec{x}-\vec{y})\rho(\vec{y})$, the Fourier image of the Green function is clearly 
$$G(\vec{k})=\frac{1}{\vec{k}^2-(a_1-a_2)\xi^2b^2\vec{k}^4}=\frac{1}{(a_1-a_2)\xi^2b^2}\left(\frac{1}{\vec{k}^2}-\frac{1}{\vec{k}^2-\frac{1}{(a_1-a_2)\xi^2b^2}}
\right)
$$
Returning to the coordinate space, we have
$$
G(\vec{x}-\vec{y})=\frac{1}{4\pi}\left(\frac{1}{|\vec{x}-\vec{y}|}-\frac{e^{-\frac{|\vec{x}-\vec{y}|}{\xi b\sqrt{a_1-a_2}}}}{|\vec{x}-\vec{y}|}\right).
$$
So, the additive, $\xi$-dependent term in the Green function is either very rapidly decaying or very rapidly oscillating at large distances. We note that since the factor $a_1-a_2$ depends on arbitrary parameters $\zeta_b$ and $\mu$,
it can be either positive or negative, thus, both extremely quick decay and extremely quick oscillation of the additive term in the Green function (and hence, in the static potential) are possible. In both cases the contributions of the additive term to physical observables would be very tiny.

\subsection{The bumblebee field self-energy}

\begin{figure}[ht] 	\includegraphics[angle=0 ,width=12cm]{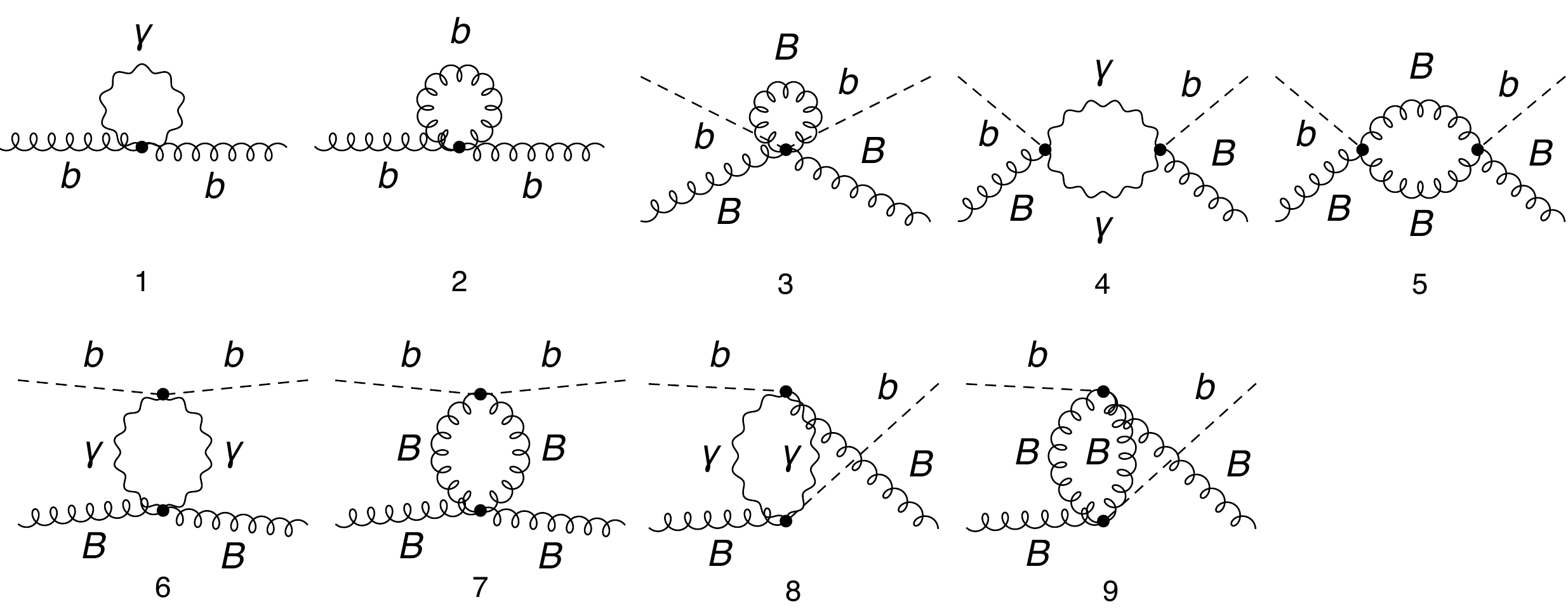} 	\caption{One-loop Bumblebee field self-energy.} 	\label{fig02} \end{figure}

Now we shall compute the self-energy of the bumblebee field, up to order $\mathcal{O}(b^2)$, corresponding to the Feynman diagram depicted in Fig. \ref{fig02}. 

First of all, let us discuss the renormalization of the bumblebee Lagrangian. We begin our calculation with writing down the effective bumblebee potential in the form
\begin{eqnarray}
{\cal L}_p=\frac{\lambda v^2}{2}B^2-\frac{\lambda}{4}(B^2)^2+O(v^4).
\end{eqnarray}
To renormalize fields and constants, we introduce the following redefinitions:
 $$
 B_{\mu}\to Z_B^{1/2}B_{\mu},\quad\, \lambda\to Z_{\lambda}\lambda, \quad\, v\to Z_{v}^{1/2}v,
 $$
 where $Z_B,Z_{\lambda},Z_{v}$ are the corresponding renormalization constants. So, our potential is renormalized as
 \begin{eqnarray}
{\cal L}_p=Z_{\lambda}Z_BZ_v\frac{\lambda v^2}{2}B^2-Z_{\lambda}Z_B^2\frac{\lambda}{4}(B^2)^2+O(v^4).
\end{eqnarray}
\noindent We note that the kinetic term can be straightforwardly renormalized with use of the wave function renormalization $Z_B$. However, the renormalization of the potential is less trivial and deserves the special discussion which we present now.

Then, let us break the Lorentz symmetry through replacement $B_{\mu}\to B_{\mu}+b_{\mu}$. We obtain
\begin{eqnarray}
{\cal L}_p&=&Z_{\lambda}Z_BZ_v\frac{\lambda v^2}{2}B^2-Z_{\lambda}Z_B^2\frac{\lambda}{4}[4(b\cdot B)^2+2b^2B^2]+\ldots =\nonumber\\
&=&\frac{\lambda v^2}{2}[Z_{\lambda}Z_BZ_v -Z_{\lambda}Z_B^2]B^2-Z_{\lambda}Z_B^2\lambda(b\cdot B)^2+\ldots,
\end{eqnarray}
where dots are for higher-order terms. 

Afterwards, we define $Z_i=1+\tilde{\delta}_i$ for each of these three renormalization constants. We arrive at
\begin{eqnarray}
{\cal L}_p&=&\frac{\lambda b^2}{2}[(1+\tilde{\delta}_{\lambda})(1+\tilde{\delta}_B)(1+\tilde{\delta}_v)-(1+\tilde{\delta}_{\lambda})(1+\tilde{\delta}_B)^2]B^2-\nonumber\\ &-&
(1+\tilde{\delta}_{\lambda})(1+\tilde{\delta}_B)^2\lambda(b\cdot B)^2=\nonumber\\
&=&\frac{\lambda b^2}{2}(\tilde{\delta}_v-\tilde{\delta}_B)B^2-
\lambda(\tilde{\delta}_{\lambda}+2\tilde{\delta}_B)(b\cdot B)^2-\lambda(b\cdot B)^2+\ldots.
\end{eqnarray}
Our key result is that while the tree-level term proportional to $B^2$, that is, the mass term for the bumblebee field, vanishes (see the discussion in the beginning of the section 3), the corresponding counterterm does not vanish. As a result, to renormalize our potential consistently, we can introduce the following counterterm Lagrangian:
\begin{eqnarray}
{\cal L}_{CT}=\frac{\delta_M}{2}B^2-\delta_{\lambda}(b\cdot B)^2,
\end{eqnarray}
where $\delta_M=\lambda v^2(\tilde{\delta}_v-\tilde{\delta}_B)$ at lowest level in $\xi$, and $\delta_{\lambda}=\lambda(\tilde{\delta}_{\lambda}+2\tilde{\delta}_B)$. Therefore, our model, considered up to the fourth order in $B_{\mu}$ is multiplicatively renormalizable, as it must be. Naturally, since we are working with a non-renormalizable theory, a set of additional operators is required to ensure that all UV divergences are absorbed, as we will see below, in line with the framework of an effective field theory for gravity.

Now, let us perform the renormalization explicitly. To start, we observe that the expressions associated with diagrams \ref{fig02}.1 and \ref{fig02}.2 are proportional to a single loop of a massless particle; hence, they result in vanishing contributions. Therefore, we need to focus on the remaining diagrams that contribute to the bumblebee field self-energy.

Next, we calculate the LV corrections to the self-energy of the bumblebee field, depicted in Fig. \ref{fig02}. The bumblebee polarization tensor up to $\mathcal{O}(\xi)$ can be expressed as
\begin{eqnarray}
	(2\pi)^D \Pi^{\mu\nu}_B(p) &=& 	 \frac{16i\pi^2\lambda \textbf{B}_0(p^2,0,0)}{3}
	\Big[4b^\mu b^\nu(7\zeta_b^2+15\zeta_b+30)+2\eta^{\mu\nu}v^2(\zeta_b^2+6\zeta_b+12) \nonumber\\
&&	+\xi \Big( \eta^{\mu\nu} (2\zeta_b (9\lambda v^4+(b\cdot p)^2)+4 (9\lambda v^4+2 v^2 p^2 -4 (b\cdot p)^2)+3\lambda v^4\zeta_b^2)\nonumber\\
&& + b^\mu b^\nu (28\lambda v^2 \zeta_b^2+\zeta_b(60 \lambda v^2 -26p^2)+60(2\lambda v^2 -p^2))\nonumber\\
&&+6(7+2\zeta_b)(b\cdot p)(b^\mu p^\nu+b^\nu p^\mu)-4 v^2 p^\mu p^\nu
	\Big)\Big]
	+\mathcal{O}(\xi^2),
\end{eqnarray}
\noindent where scaleless integrals are made vanishing.

After returning to the coordinate representation, we can express the corresponding one-loop quadratic part of the bumblebee effective Lagrangian as
\begin{eqnarray}
\mathcal{L}_{eff}^{(1)} &=&-\left(\delta_{b_1}- \frac{64i \pi ^2 \lambda\xi}{3(2\pi)^D}  \textbf{B}_0\left(p^2,0,0\right)\right) v^2 B^{\mu\nu} B_{\mu\nu} 
\nonumber\\
&&+\frac{1}{2}\left(\delta_{b_2}-\frac{16 i \pi ^2 \lambda\xi  \left(7+ \zeta_b\right)}{(2\pi)^D}\textbf{B}_0\left(p^2,0,0\right) \right) b^\alpha B_{\mu\alpha} b_{\nu} B^{\mu\nu}\nonumber\\
&&+\left(\delta_{b_3}+ \frac{16i\pi ^2 \lambda  \xi }{3(2\pi)^D} \textbf{B}_0\left(p^2,0,0\right)\right) b^2(\partial^\mu B_\mu)^2
\nonumber\\
&&+\left(\delta_{b_4}+\frac{8i \pi^2\lambda  \xi  \left(13+7\zeta_b\right) }{3(2\pi)^D}\textbf{B}_0\left(p^2,0,0\right)  \right) b^\mu (\partial_\mu B_\alpha) b^\nu (\partial_\nu B^\alpha)\nonumber\\
&&+\left(\delta_{b_5}-
\frac{8 i \pi^2 \lambda  \xi  \left(9+7\zeta_b \right)}{(2\pi)^D} \textbf{B}_0\left(p^2,0,0\right)
\right) b^\mu (\partial_\alpha B_\mu) b^\nu (\partial^\alpha B_\nu)
\nonumber\\
&&+\frac{1}{2}\left(\delta_{M^2}-\frac{16 i \pi^2 \lambda^2 v^2 \left(2+ 3 \xi  v^2\right)(6+3\zeta_b+2\zeta_b^2)}{(2\pi)^D} \textbf{B}_0\left(p^2,0,0\right) \right) B^2
\nonumber\\
&&-\frac{\lambda}{2}\left(\delta_\lambda- \frac{16 i \pi ^2 \lambda  \left(1+\xi  v^2\right)(30+15\zeta_b+7\zeta_b^2)}{3(2\pi)^D} \textbf{B}_0\left(p^2,0,0\right) \right)\left[b^2 B^2+2 (b^\mu B_\mu)^2\right]\nonumber\\
&&+\mathcal{O}(\xi^2),
\end{eqnarray}
\noindent from which it is straightforward to obtain the counterterms by imposing finiteness in the MS scheme of renormalization.

We find that our divergent contribution involves, first, the Maxwell-like term; second, the aether-like term; third, several additional derivative-dependent terms; and fourth, both Lorentz-invariant and Lorentz-violating mass terms, similar to those presented in \cite{ourbumb2}. Furthermore, we also identify the finite nonlocal counterparts of these operators.


\section{Summary}


We formulated a coupling of the electromagnetic field to the metric-affine bumblebee gravity. Then, we restricted ourselves to the case of the weak field regime where the spacetime metric is expanded up to the second-order in the non-minimal coupling $\xi$ around the flat background. Furthermore, the new effective gauge-bumblebee theory arises when the action is expanded up to the second order in the non-minimal coupling, $\xi$. In this scenario, we considered small fluctuations of the bumblebee field around one of the vacua introducing spontaneous Lorentz symmetry breaking and calculated the lower quantum corrections in the resulting theory. We emphasized that, as a result of LSB, such an effective theory yields some unconventional couplings between the bumblebee and the electromagnetic fields, and self-couplings of the bumblebee field. In the one-loop approximation, we have generated non-local contributions to the Podolsky-like  and aether-like LV terms which have not be obtained in \cite{ourbumb2}. Also, we found the lower (quadratic) quantum contribution to the effective action of the bumblebee field up to the second order in derivatives.

Let us briefly discuss the physical significance of our results. We note that the bumblebee field, within our framework, represents itself as an effective source for the non-metricity tensor in the weak field approximation, cf. Eqs. (\ref{gexpand},\ref{gexpand1}). Therefore, we presented a mechanism allowing to obtain the LV contributions coming from the post-Minkowskian corrections to the effective action of the electromagnetic field, together with the lower LV contribution to the effective gravity action within the bumblebee model, thus, opening the way for dynamical generation of various terms listed in \cite{KosLiGrav}. We expect that these additive terms can be used for studying various LV effects for the weak gravity, in particular, for different aspects of its interaction with the gauge fields.

Regarding other possible gauge-vector couplings, we note that certainly, our theory is not unique one allowing for coupling of bumblebee and gauge fields. We can consider also the model like
\begin{eqnarray}
{\cal L}&=&-\frac{1}{4}B^{\mu\nu}B_{\mu\nu}-V(B^{\mu}B_{\mu}\pm b^2)-\nonumber\\
&-&\frac{1}{4}\chi_1(B^{\mu}B_{\mu}\pm b^2)F_{\mu\nu}F^{\mu\nu}-\frac{1}{4}\chi_2(B^{\mu}B_{\mu}\pm b^2)F_{\mu\nu}\tilde{F}^{\mu\nu},
\end{eqnarray}
where $\tilde{F}^{\mu\nu}=\frac{1}{2}\epsilon^{\mu\nu\lambda\rho}F_{\lambda\rho}$ is the dual tensor, and $\chi_1$, $\chi_2$ are some functions assumed to be of the form $\chi_{1,2}(z)=1+z+\ldots$. Such a theory can be treated as a reminiscence of effective models for field dependent magnetic permeability \cite{perm}. However, unlike \cite{perm} where the permeability was described by a scalar field, here it is presented with a vector one which probably can allow for studies of anisotropic manifestations of magnetisation. We expect to study such theories in forthcoming papers.

\textbf{Acknowledgments.} 
This work was partially supported by Conselho Nacional de Desenvolvimento Cient\'{i}fico e Tecnol\'{o}gico (CNPq). The work by A. Yu. P. has been partially supported by the CNPq project No. 303777/2023-0. The work of A. C. L. has been partially supported by the CNPq project No. 404310/2023-0. P. J. Porf\'{\i}rio would like to acknowledge the Brazilian agency CNPq, grant No. 307628/2022-1.

{\bf Data Availability Statement:} No Data associated in the manuscript.

{\bf Conflict of interest statement:} The authors declare that they have no conflict of interests.
 
 \appendix
 
 \section*{Integrals notation}
 
We are using the same notations and conventions for Passarino-Veltman one-loop integrals as in Ref. \cite{Mertig:1990an}. The integrals involved in our calculations are

\begin{eqnarray}
	\textbf{B}_0(p^2,0,0)&=&\frac{1}{i\pi^2}\int{d^Dk}\frac{1}{(k+p)^2(k^2)}=\frac{1}{\epsilon}-\gamma+2-\ln\left(-\frac{p^2}{\pi\mu^2}\right);\\
	\textbf{B}_0(0,0,0)&=&\frac{1}{i\pi^2}\int{d^Dk}\frac{1}{(k^2)^2}=0,
\end{eqnarray}
 
\noindent where  $\mu$ is a mass scale introduced by the regularization, $\epsilon=(4-D)/2$ is dimensional regularization regulator and $\gamma$ denotes the Euler-Mascheroni constant. Notice that we adopt the prescription where all scaleless integrals vanish \cite{Abreu:2022mfk}. 

 
 The integral  $\textbf{C}_0$ is defined as 
   \begin{eqnarray}
  	\textbf{C}_0(0,p^2,p^2,0,0,0)&=&\frac{1}{i\pi^2}\int{d^Dk}\frac{1}{(k+p)^2(k^2)^2}=-\frac{\textbf{B}_0(p^2,0,0)}{p^2},
  \end{eqnarray}
 \noindent where the final equality is derived using Tarasov's algorithmic approach \cite{Tarasov:1997kx}.


\end{document}